\documentclass[multphys,vecphys]{svmult}

\usepackage{makeidx}     
\usepackage{graphicx}    
\usepackage{multicol}    

\makeindex             

\begin{document}
\title*{A Masing Molecular Cloud in the Central Parsecs of Mrk~348}
\titlerunning{Maser in Mrk348}
\author{Peck, A. B.\inst{1} \and Henkel, C.\inst{2} \and Ulvestad, J. S.\inst{3}}
\institute{Harvard Smithsonian CfA, SMA Project, P.O. Box 824, Hilo,
  HI 96720 USA
\texttt{apeck@cfa.harvard.edu}
\and Max-Planck-Institut f\"ur Radioastronomie, Auf dem H\"ugel 69,
  53121 Bonn, Germany \texttt{p220hen@mpifr-bonn.mpg.de} \and NRAO,
  P.O. Box 0, Socorro, NM 87801 USA \texttt{julvesta@nrao.edu}}
%
%
\maketitle

We report new observations of the H$_2$O megamaser in the Seyfert 2
galaxy Mrk~348.  The line is redshifted by $\sim$130 km s$^{-1}$ with
respect to the systemic velocity, is extremely broad, with a FWHM of
130 km s$^{-1}$, and has no detectable high velocity components within
1500 km s$^{-1}$ on either side of the observed line.  The unusual
line profile led us to suspect that this source, like NGC~1052
\cite{cla98}, might belong to a class of megamaser galaxies in which
the amplified emission is the result of an interaction between the
radio jet and an encroaching molecular cloud, rather than occurring in
a circumnuclear disk.  Our initial VLBA observations show that the
maser emission emanates entirely from a region $\le$0.25 pc in extent,
located toward a continuum component thought to be associated with the
receding jet \cite{pec03}.  The very high linewidth occurring on such
small spatial scales and the rapid variability indicate that the
H$_2$O emission is more likely to arise from a shocked region at the
interface between the energetic jet material and the molecular gas in
the cloud where the jet is boring through, than simply as the result
of amplification by molecular clouds along the line of sight to the
continuum jet.  The orientation of the radio jets close to the plane
of the sky also results in shocks with the preferred orientation for
strong masers from our vantage point.  Single-dish monitoring with the
Effelsberg 100m telescope showed that the line and continuum emission
``flared'' on very similar timescales.  The close temporal correlation
between this activity in the maser emission and the continuum flare
further suggest that the masing region and the continuum hotspots are
nearly equidistant from the central engine and may be different
manifestations of the same dynamical events.  The study of this newly
discovered type of H$_2$O megamaser can provide detailed information
about the conditions in the ISM in the central 1-10 pc of active
galactic nuclei.

\section{Introduction}
\label{sec:1}

H$_2$O\ megamasers are best known as a means to probe the accretion
disks in Seyfert galaxies. In the most famous source, NGC~4258, a
thin, slightly warped, nearly edge-on disk orbits in Keplerian
rotation around a central mass of 4$\times$10$^{7}$
{\hbox{M$_{\odot}$}}\ (e.g. \cite{miy95}).  VLBI studies have been
used to determine the size and shape of this warped molecular disk as
traced by the maser spots.  A few other sources show evidence of a
toroidal structure, but the distribution of maser spots is not as well
understood.  There is now evidence, however, for a distinct class of
H$_2$O megamaser. In these sources the amplified emission is the
result of an interaction between the radio jet and an encroaching
molecular cloud, rather than occurring in a circumnuclear disk.  The
only known sources in this class were NGC~1068 \cite{gal96} and the
Circinus galaxy \cite{gre01}, which appear to have both a
circumnuclear disk and maser emission arising along the edges of an
ionization cone or outflow, and NGC~1052, in which the masers appear
to arise along the jet and have a full width at half maximum (FWHM)
$\sim$90 km s$^{-1}$ \cite{cla98}.  We have recently identified the
fourth such source, Mrk~348, a Seyfert 2 galaxy with a low inclination
angle and an exceptionally bright and highly variable nuclear radio
source \cite{fal00}.  Observations of Mrk~348 using Global VLBI at
4.8~GHz \cite{roy99} show two faint continuum hotspots on either side
of a much brighter central peak.  The relative intensities of the two
outer components are quite similar, indicating that relativistic
beaming effects are probably minimal and the jet axis should be close
to the plane of the sky.  Ground-based observations \cite{sim96} show
evidence of an ionization cone with a half-angle of
$\sim$45{$^\circ$}, which also suggests a jet axis fairly close to the
plane of the sky.

The initial detection of the flaring maser in Mrk~348 using the
Effelsberg 100m telescope took place in 2000 March \cite{fal00}. The
H$_2$O maser line in Mrk~348 is extremely broad, with a FWHM of
$\sim$130 km s$^{-1}$, though in many of the monitoring epochs the
emission appears to consist of 2 lines which can be tentatively fit
with a broad Gaussian function at $\sim$4609 km s$^{-1}$ with
FWHM$\sim$100 km s$^{-1}$ and a narrower one at $\sim$4678 km s$^{-1}$
with FWHM$\sim$60 km s$^{-1}$.  The amplitudes of each component vary
significantly on very short timescales.  There are no detectable high
velocity components within 1500 km s$^{-1}$ on either side of the
strong emission line \cite{fal00}.

\begin{figure}
\centering
\vspace{6cm}
\includegraphics{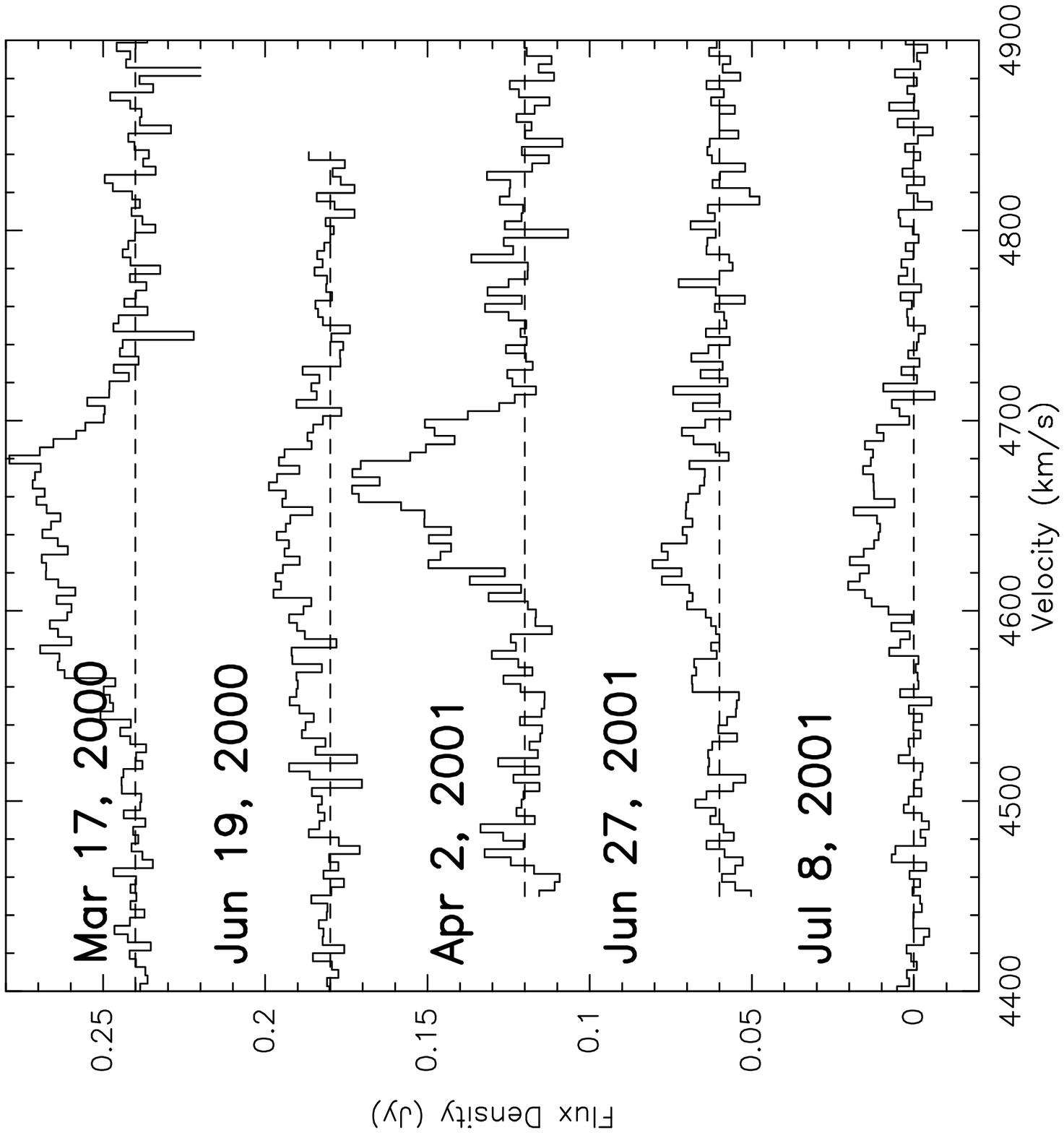}
\includegraphics{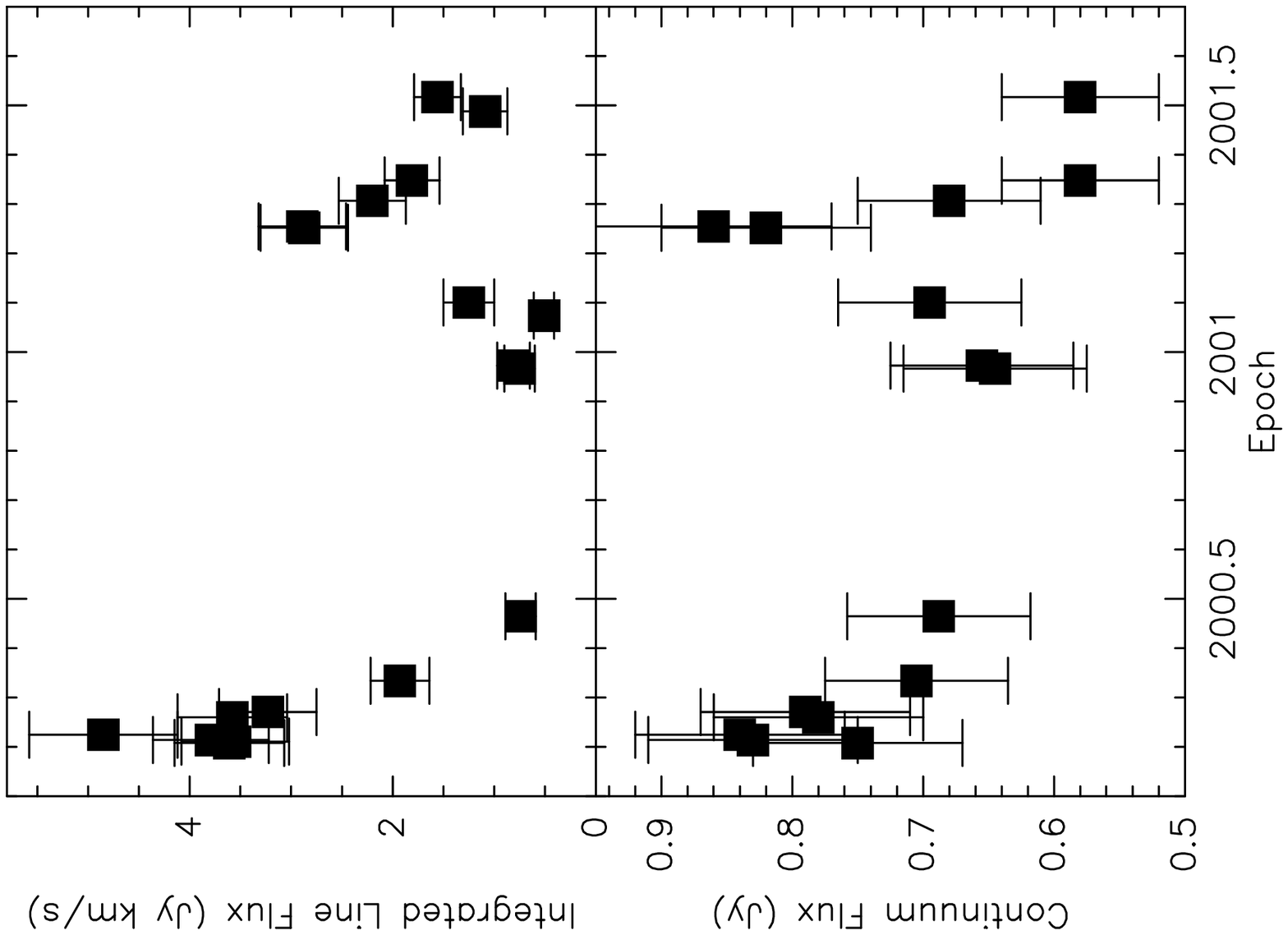}
\caption{{\bf Left)}5 H$_2$O spectra from the Effelsberg 100-m telescope
  chosen to illustrate the variations in the profiles seen on
  timescales of days or months. {\bf Top right)} Integrated line flux
  density vs. time; {\bf Bottom right)} continuum flux density vs. time
  measured with the Effelsberg 100-m telescope.  The variations in
  line and continuum intensity can be seen to vary concurrently.}
\label{fig:peckfig1}       
\end{figure}

Figure~\ref{fig:peckfig1}a shows 5 spectra which were chosen to emphasize
the significant variations in the line profiles over our first 16
months of observations.  The first profile is the discovery spectrum,
where both components had a flux of around 30~mJy.  In the second
profile, taken within a week of our VLBA observations, both components
verged on undetectability and then, in the third profile, the higher
velocity component flared to twice the intensity of the year previous,
and the lower velocity component disappeared altogether.  This change
was reversed in June, shown in the fourth profile, when the lower
velocity component was present, though the intensity was low, and the
higher velocity component vanished.  Within two weeks, both components
were once again nearly equal.
Figure~\ref{fig:peckfig1}b shows the variation in line and continuum
flux density with respect to time.  Comparison shows that the
continuum flux varies loosely with the maser flux.  The close temporal
correlation between the flaring activity in the maser emission and the
continuum flare suggest that the masing region and the continuum
hotspots are nearly equidistant from the central engine and may be
different manifestations of the same dynamical events. VLBA continuum
observations indicate that the center of activity in Mrk~348\ is
likely to be located between the two strongest central components
\cite{pec03}.  We hypothesize that at an epoch prior to 1998.75, jet
components were ejected simultaneously in the approaching and receding
jets, leading to both the continuum flare and the new flaring activity
of the maser.

\begin{figure}
\centering
\vspace{5.5cm} 
\includegraphics{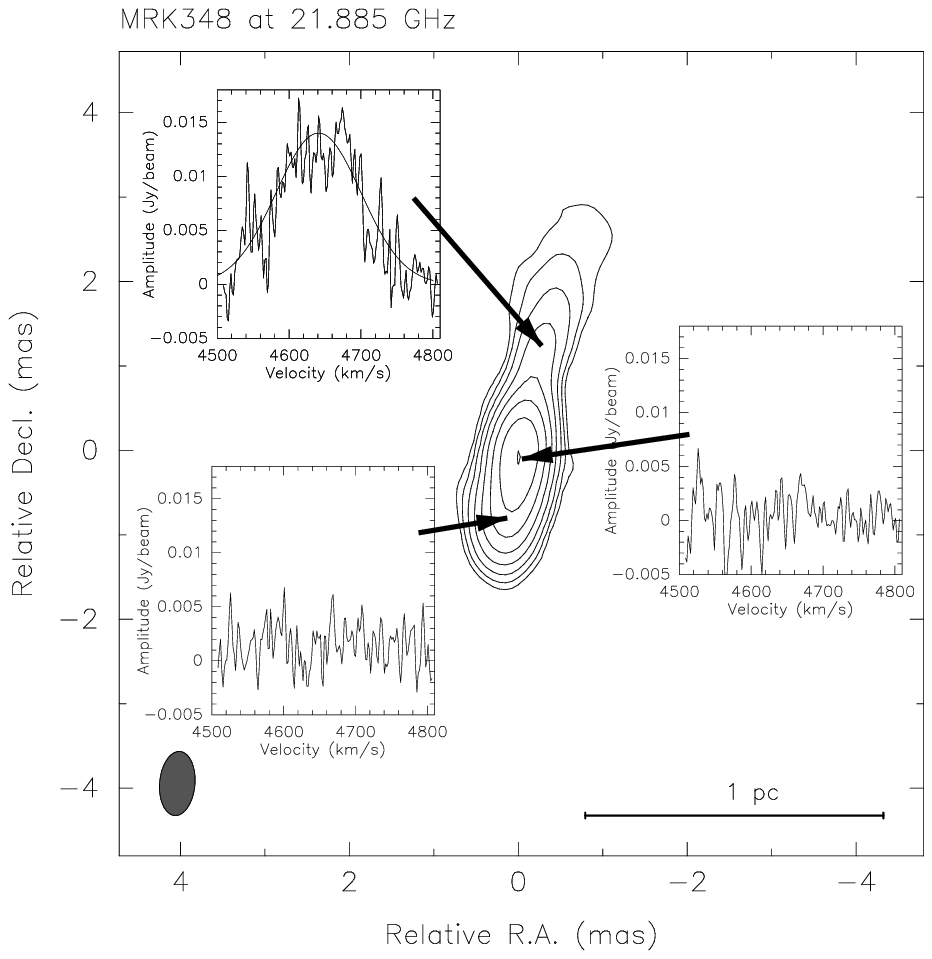}
\includegraphics{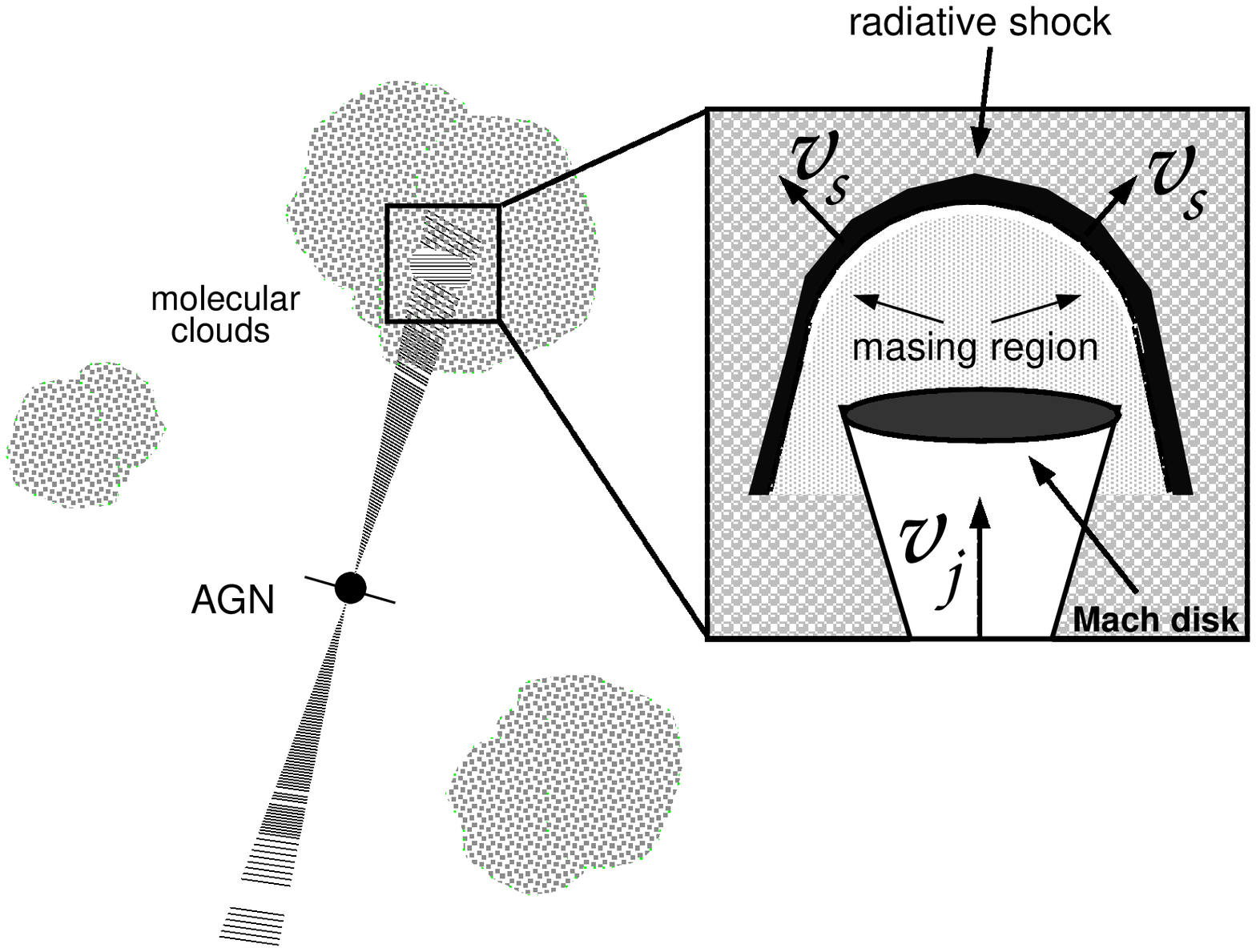}
\caption{{\bf Left)}H$_2$O line profiles toward Mrk~348, superimposed
on the continuum map made from 20 line-free channels extracted from
the low frequency end of the observed frequency range.  The RMS noise
in the line profiles is $\sim$4~mJy~beam$^{-1}$~channel$^{-1}$.  The
continuum image is naturally weighted and the lowest contour is
5~mJy. The RMS noise in the continuum map is less than
1~mJy~beam$^{-1}$.  {\bf Right)}A cartoon model of the expanding
bubble caused by the jet material impacting the molecular cloud.  The
maser emission arises within the region surrounded by the radiative
shock.}
\label{fig:peckfig2}       
\end{figure}

Figure~\ref{fig:peckfig2}a shows three integrated line profiles toward
Mrk~348 taken using the VLBA.  The maser emission is clearly seen to
lie along the line of sight to the fainter northern continuum
component, rather than the brightest region of the continuum source
which is thought to contain the core.  The Gaussian fit to the line
shown in the upper left profile has an amplitude of 14{$\pm$}2~mJy and
an integrated flux of 2.11{$\pm$}0.34 Jy/beam/km s$^{-1}$, indicating
that all of the flux measured in the Effelsberg May 2 observation has
been recovered.  The FWHM is 139{$\pm$}11 km s$^{-1}$ centered on
4641{$\pm$}2.2 km s$^{-1}$, consistent with the single-dish
measurements and redshifted by 134 km s$^{-1}$ with respect to the
systemic velocity.

In order to produce the maser emission detected in this source, the
gas in the molecular cloud within the central parsecs of Mrk~348 must
have a pre-shock density ranging from around 10$^6$ cm$^{-3}$ to a few
times 10$^9$ cm$^{-3}$.  An expanding bow-shock being driven into this
cloud by the AGN jet has a velocity between 135 km s$^{-1}$ and 0.5c
in the direction of jet propagation, and between 35 km s$^{-1}$ and
300 km s$^{-1}$ at various points along the oblique edges (Figure
\ref{fig:peckfig2}b).  This shock generates a region of very high
temperature, ($\le$10$^5$ K), which dissociates the molecular gas and
to some extent shatters the dust grains expected to be present and/or
evaporates their icy mantles.  Immediately following this shock, H$_2$
begins forming on the surviving dust grains when the temperature has
dropped to $\sim$1000 K, and this in turn provides sufficient heating
to stabilize the temperature at $\sim$400 K, with gas densities of
$\sim$10$^8$ cm$^{-3}$ and ultimately an H$_2$O\ abundance as high as
10$^{-5}$.  Ammonia is not found because the levels of UV radiation
remain too high, but the detection of other molecular species will
help to test this scenario.

%
%
%
%

%
%



\printindex
\end{document}